\documentclass[a4paper,aps,prd,10pt,preprintnumbers,showpacs,twocolumn,superscriptaddress,nofootinbib,amsmath,amssymb]{revtex4-1}
\usepackage[utf8]{inputenc}
\usepackage[T1]{fontenc}
\usepackage{cmap}

\def\m{\overline{m}}
\def\a{\widetilde{\alpha}}
\def\M{{\cal M}}
\def\Q{{\cal Q}}
\def\LE{\overline{\Lambda}}
\def\QE{\overline{\cal Q}}
\def\O{{\cal O}}

\begin{document}
\title{BTZ black holes with higher curvature corrections in the $3D$ Einstein-Lovelock gravity}

\author{R. A. Konoplya}\email{roman.konoplya@gmail.com}
\affiliation{Research Centre for Theoretical Physics and Astrophysics, Institute of Physics, Silesian University in Opava, Bezručovo nám. 13, CZ-74601 Opava, Czech Republic}
\affiliation{Peoples Friendship University of Russia (RUDN University), 6 Miklukho-Maklaya Street, Moscow 117198, Russian Federation}

\author{A. Zhidenko}\email{olexandr.zhydenko@ufabc.edu.br}
\affiliation{Research Centre for Theoretical Physics and Astrophysics, Institute of Physics, Silesian University in Opava, Bezručovo nám. 13, CZ-74601 Opava, Czech Republic}
\affiliation{Centro de Matemática, Computação e Cognição (CMCC), Universidade Federal do ABC (UFABC), \\ Rua Abolição, CEP: 09210-180, Santo André, SP, Brazil}

\begin{abstract}
The regularization procedure for getting the four-dimensional nontrivial Einstein-Gauss-Bonnet effective description of gravity and its Lovelock generalization has been recently developed. Here we propose the regularization for the three-dimensional gravity, which is based on the rescaling of the coupling constants and, afterward, taking the limit $D \to 3$. We obtain the generalization of the Bañados-Teitelboim-Zanelli solution in the presence of the higher curvature (Gauss-Bonnet and Lovelock) corrections of any order. The obtained general solution shows a peculiar behavior: The event horizon is allowed not only for asymptotically anti-de Sitter spacetimes, but also for the de-Sitter and flat cases, when the Gauss-Bonnet coupling constant is negative. The factor of the electric charge is analyzed as well for various branches of the solution and the Hawking temperature is obtained.
\end{abstract}

\maketitle

\section{Introduction}
Black holes in theories of gravity of lower than four dimensions play an important role for our understanding of properties of black holes as well as strongly coupled dual systems  \cite{Banados:1992wn,Birmingham:2001pj,Grumiller:2002nm,Ge:2017fix}. Analysis of various phenomena in the background of the lower-dimensional black holes are sometimes remarkably simple in comparison with the full higher-dimensional problem and, frequently, allows for an analytic solution. Probably the most successful black-hole solution of this kind is the $(2+1)$-dimensional asymptotically anti-de Sitter (AdS) black hole called the Bañados-Teitelboim-Zanelli (BTZ) solution \cite{Banados:1992wn}. Various properties of the BTZ black hole and its generalizations in modified theories of gravity were considered (see \cite{Cardoso:2001hn,Konoplya:2004ik,Eune:2013qs,Shu:2014eza,Gupta:2015uga,Hendi:2016pvx,Gupta:2017lwk,Stikonas:2018ane,Hirano:2019ugo,Cong:2019bud} and references therein). Within the three-dimensional Einstein-Maxwell theory the black-hole solutions exist only in the presence of a negative cosmological constant, that is, only asymptotically anti-de Sitter black holes are allowed.

Recently, an interesting formulation of the four-dimensional Einstein-Gauss-Bonnet gravity was suggested in \cite{Glavan:2019inb}, which has diffeomorphism invariance and second-order equations of motion. It was claimed that the approach bypasses the Lovelock theorem \cite{Lovelock}, which means that the constructed gravity is different from the pure Einstein theory in $(3+1)$-dimensional spacetime.

The approach is first formulated in $D > 4$ dimensions, and then the four-dimensional theory is defined as the limit $D \to 4$ of the higher-dimensional theory after the rescaling of the coupling constant. Notice, that the same regularized four-dimensional black-hole metrics with Gauss-Bonnet (GB) corrections were obtained earlier within different approaches \cite{Tomozawa:2011gp,Cognola,Cai:2009ua}, such as, for example, looking for quantum corrections to the black-hole entropy. The properties of black holes obtained within this approach, such as (in)stability, quasinormal modes and shadows, were considered in \cite{Konoplya:2020bxa}, while the Hawking radiation analyzed in \cite{Konoplya:2020cbv,Zhang:2020qam}. The innermost circular orbits were analyzed in \cite{Guo:2020zmf}. The generalization to the charged black holes and asymptotically anti-de Sitter and de Sitter cases in the $4D$ Einstein-Gauss-Bonnet gravity was considered in \cite{Fernandes:2020rpa} and to the higher curvature corrections in \cite{Konoplya:2020qqh,Casalino:2020kbt}. Some further properties of black holes and stars for this novel formulation of gravity, such as axial symmetry, thermodynamics and others, were considered in \cite{Wei:2020ght,Kumar:2020owy,Hegde:2020xlv,Ghosh:2020vpc,Doneva:2020ped,Zhang:2020qew}.

It is important to notice here that such a formulation, apparently, cannot be performed in four dimensions in a consistent manner. Because of the Lovelock theorem, the four-dimensional theory cannot be obtained from the action principle and equations for the coupled fields in their general covariant form have no proper limit for $D\to4$. In particular, the contribution from the Gauss-Bonnet term does not obey the Bianchi identity, leading to an inconsistency when coupling of conserved fields in the four-dimensional theory \cite{Gurses:2020ofy}. In order to solve the above problem, it has been proposed to perform the regularization by introducing explicitly a conformal metric factor in the $D$-dimensional action, which becomes an additional scalar field in the limit $D\to4$ \cite{Fernandes:2020nbq}. In two dimensions this method leads to the equations which are equivalent to the regularized Einstein theory \cite{Mann:1992ar}. When applied to the limit $D\to4$ of the Einstein-Gauss-Bonnet theory, it admits the same cosmological and black-hole solutions \cite{Kobayashi:2020wqy,Lu:2020iav,Hennigar:2020lsl}, yet, it is not possible to prove full equivalence of the theories, obtained within these two approaches, and the scalar field may carry an additional hidden degree of freedom \cite{Fernandes:2020nbq}.

Although the obtained here lower-dimensional black-hole metrics could be used as a viable background for test fields, when the matter field is not simply propagating in the background of a spacetime given by some metric but is strongly coupled to gravity, the $D$-dimensional treatment of a problem is necessary with the subsequent rescaling of the coupling constant leading to the regularization in the $D\to4$ limit.  Thus, when the coupling to gravity comes into play, for example under the analysis of gravitational perturbations and stability of black holes, then the perturbation equations must first be considered in $D$-dimensional case and then the dimensional regularization must be performed for the whole system, as was done in \cite{Konoplya:2020bxa,Konoplya:2020juj}.

Here, following the approach of \cite{Glavan:2019inb}, we suggest a similar regularization for the $(2+1)$-dimensional gravity allowing for an electric charge, cosmological constant (both positive and negative), and Lovelock terms of any order. We find the generalization of the $(2+1)$-dimensional charged BTZ black hole in this approach, which includes higher curvature (Gauss-Bonnet and Lovelock) corrections.

We will show that the black-hole metric has the following simple form for the $3D$-Einstein-Gauss-Bonnet gravity:
$$f(r)=1-\frac{r^2}{2\a_2}\left(-1\pm\sqrt{1+4\a_2\frac{\Lambda(r^2-r_H^2)+\dfrac{\a_2}{r_H^2}+1}{r^2}}\right),$$
where $\a_2$ is the GB coupling constant, $\Lambda$ is the cosmological constant, and $r_{H}$ is the radius of the event horizon, such that for the sign ``+'' corresponds to the branch, perturbative in $\a_2$, for which $1+2\a_2/r_H^2>0$, while the sign ``-'' appears for the nonperturbative branch and implies that $1+2\a_2/r_H^2<0$. This simple metric is further generalized to the case of a charged black hole and black string, as well as to the higher-order Lovelock terms.
Unlike the classical BTZ metric, our general solution allows for an event horizon even for asymptotically de Sitter or flat black holes and strings, provided the GB coupling constant is negative. We consider some basic properties of these perturbative and nonperturbative solutions. The Hawking temperature and the horizon structure are discussed for these cases.

Our paper is organized as follows. In Sec.~\ref{sec:blackhole} we briefly describe the generic static maximally symmetric solution in the $3D$ Einstein-Lovelock gravity, allowing for an electric charge and the $\Lambda$-term. In Sec.~\ref{sec:horizon} we go over to the units of the radius of the event horizon, consider the basic properties of the general black-hole solution, and calculate the Hawking temperature. In Sec. \ref{sec:GaussBonnet} we discuss different branches of the $3D$-Einstein-Gauss-Bonnet solution and show that, in addition to the nonperturbative branch, there is the perturbative branch which is a generalization of the charged BTZ solution. In Sec.~\ref{sec:Lovelock} we consider the general solution for a particular case of the third-order Einstein-Lovelock gravity. Finally, in Conclusions, we summarize the obtained results and discuss some open questions.

\section{Static solutions in the three-dimensional Lovelock gravity}\label{sec:blackhole}
The Lagrangian density of the Lovelock-Maxwell theory has the form \cite{Lovelock,Kofinas:2007ns}
\begin{eqnarray}\label{Lagrangian}
  &&\mathcal{L} = -2\Lambda+\frac{1}{4}F^{\mu\nu}F_{\mu\nu}
  \\\nonumber&&
  +\sum_{m=1}^{\m}\frac{1}{2^m}\frac{\alpha_m}{m}
  \delta^{\mu_1\nu_1\ldots\mu_m\nu_m}_{\lambda_1\sigma_1\ldots\lambda_m\sigma_m}\,
  R_{\mu_1\nu_1}^{\phantom{\mu_1\nu_1}\lambda_1\sigma_1}\ldots R_{\mu_m\nu_m}^{\phantom{\mu_m\nu_m}\lambda_m\sigma_m},
\end{eqnarray}
where
$$\delta^{\mu_1\mu_2\ldots\mu_p}_{\nu_1\nu_2\ldots\nu_p}=\det\left(
\begin{array}{cccc}
\delta^{\mu_1}_{\nu_1} & \delta^{\mu_1}_{\nu_2} & \cdots & \delta^{\mu_1}_{\nu_p} \\
\delta^{\mu_2}_{\nu_1} & \delta^{\mu_2}_{\nu_2} & \cdots & \delta^{\mu_2}_{\nu_p} \\
\vdots & \vdots & \ddots & \vdots \\
\delta^{\mu_p}_{\nu_1} & \delta^{\mu_p}_{\nu_2} & \cdots & \delta^{\mu_p}_{\nu_p}
\end{array}
\right)$$
is the generalized totally antisymmetric Kronecker delta, $R_{\mu\nu}^{\phantom{{\mu\nu}}\lambda\sigma}$ is the Riemann tensor, $\alpha_1=1/8\pi G=1$ and $\alpha_2,\alpha_3,\alpha_4,\ldots$ are arbitrary constants of the theory.

The Euler-Lagrange equations, corresponding to the Lagrangian density (\ref{Lagrangian}), read \cite{Takahashi:2012np}
\begin{eqnarray}\nonumber&&
  \Lambda\delta^{\mu}_{\nu}-\sum_{m=1}^{\m}\frac{\alpha_m}{2^{m+1}m}
  \delta^{\mu\mu_1\nu_1\ldots\mu_m\nu_m}_{\nu\lambda_1\sigma_1\ldots\lambda_m\sigma_m}
R_{\mu_1\nu_1}^{\phantom{\mu_1\nu_1}\lambda_1\sigma_1} \ldots R_{\mu_m\nu_m}^{\phantom{\mu_m\nu_m}\lambda_m\sigma_m}
\\\label{Lovelock-Maxwell}
&&\phantom{F^{\mu\nu}_{~;\nu}}=\frac{1}{2}F^{\mu\sigma}F_{\nu\sigma}-\frac{1}{8}F^{\lambda\sigma}F_{\lambda\sigma}\delta^{\mu}_{\nu}\,,
\\\nonumber&&F^{\mu\nu}_{~;\nu}=\frac{1}{\sqrt{-g}}\partial_{\nu}\sqrt{-g}F^{\mu\nu}=0\,.
\end{eqnarray}

The antisymmetric tensor in (\ref{Lovelock-Maxwell}) is nonzero only when the indices $\mu,\mu_1,\nu_1,\mu_2,\nu_2,\ldots\mu_m,\nu_m$ are all distinct. Notice that for $D=2\m$ the $\m$-order correction in the Lagrangian (\ref{Lagrangian}) does not vanish. Nevertheless it does not contribute to the equations of motion, because it is a topological invariant which leads to a surface integral in the action. Thus, the general Lovelock theory is such that $2\m <D$. In particular, for $D=4$ and $D=3$, we have $\m=1$, which corresponds to the Einstein theory \cite{Lovelock}.

We consider the general static solution, described by the metric
\begin{eqnarray}\label{Lmetric}
ds^2&=&-f(r)dt^2+\frac{1}{f(r)}dr^2 + r^2d\Omega_n^2,
\\\nonumber
f(r)&=&\kappa-r^2\psi(r)\,,
\end{eqnarray}
where $d\Omega_n^2$ is a $(n=D-2)$-dimensional space with a constant curvature $\kappa=-1,0,1$ and the only nonzero components of the electromagnetic strength tensor are
\begin{equation}\label{Echarge}
F^{tr}=-F^{rt}=E(r).
\end{equation}

Then equations (\ref{Lovelock-Maxwell}) can be reduced to the following form \cite{Takahashi:2012np}
\begin{eqnarray}\label{EEq}
\frac{d}{dr}r^{D-2}E(r) &=& 0\,, \\
\label{PEq}
-\frac{D-2}{2r^{D-2}}\frac{d}{dr}r^{D-1}\left(P[\psi(r)]-\lambda\right) &=& -\frac{E(r)^2}{4}\,,\\
\label{AEq}
-\frac{1}{2r^{D-3}}\frac{d^2}{dr^2}r^{D-1}\left(P[\psi(r)]-\lambda\right) &=&  \frac{E(r)^2}{4}\,,
\end{eqnarray}
where
$$\lambda=\frac{2\Lambda}{(D-1)(D-2)}$$
and the function $P[\psi]$ is defined as follows:
\begin{eqnarray}\label{Ppsi}
P[\psi]&=&\psi+\sum_{m=2}^{\m}\frac{\alpha_m}{m}\frac{(D-3)!}{(D-2m-1)!}\psi^m
\\\nonumber
&=&\psi+\sum_{m=2}^{\m}\a_m\psi^m.
\end{eqnarray}
Notice that equation~(\ref{AEq}) is not independent and follows from (\ref{EEq}) and (\ref{PEq}).

The new constants $\a_m$ are introduced as in \cite{Konoplya:2017lhs}:
\begin{equation}\label{amdef}
\a_m=\frac{\alpha_m}{m}\frac{(D-3)!}{(D-2m-1)!}=\frac{\alpha_m}{m}\prod_{p=1}^{2m-2}(D-2-p)\,.
\end{equation}
Considering finite values of $\a_m$, one can see that (\ref{PEq}) and (\ref{AEq}) are finite for any $D\geq3$. In this way we can perform dimensional regularization of the generic static solution (\ref{Lmetric}) in the Einstein-Lovelock theory for $D\leq2\m$.

By integrating (\ref{EEq}) we find that
\begin{equation}\label{EQ}
E(r)=\frac{\Q}{r^{D-2}}\,,
\end{equation}
where the integration constant $\Q$ is the electric charge.

After integration of (\ref{PEq}), we obtain the algebraic equation for $\psi(r)$:
\begin{equation}\label{MEq}
P[\psi(r)]=\frac{2\M}{r^{D-1}}-\frac{Q^2}{r^{2(D-2)}}+\frac{2\Lambda}{(D-1)(D-2)}\,,
\end{equation}
where the arbitrary constant $\M$ defines the asymptotic mass \cite{Myers:1988ze}:
\begin{equation}\label{MADM}
M=\frac{(D-2)\pi^{D/2-3/2}}{4\Gamma(D/2-1/2)}\M,
\end{equation}
and
\begin{equation}\label{charge}
Q=\Q\sqrt{\frac{2}{(D-2)(D-3)}}.
\end{equation}

When considering the limit $D\to3$, one can use two approaches to deal with the electromagnetic field. One approach, inferred in \cite{Konoplya:2020qqh}, is to take the limit $\Q\to0$ as $D\to3$ such that $Q$ remains finite. In this case, equation~(\ref{MEq}) for $\psi(r)$ reads
\begin{equation}\label{MEq1}
P[\psi(r)]=\frac{2\M-Q^2}{r^2}+\Lambda=\frac{8M-Q^2}{r^2}+\Lambda\,.
\end{equation}
Although the electric charge vanishes, the quantity $Q$ leads to an additional term, which is subtracted from the asymptotic mass in the metric function. As a result the constant (effective mass) is not necessary positive. Within this regularization of the higher-dimensional electrodynamics, the three-dimensional electrodynamics leads to the redefinition of mass and is, thereby, trivial.

A more natural approach is first to formulate the regularization of the gravitational sector and then to impose a three-dimensional electromagnetic field. In this way we consider  the solution to (\ref{PEq}) for a finite value of $\a_m$ and $\Q$, which reads
\begin{equation}\label{MEq2}
P[\psi(r)]=\frac{\Q^2\ln(r/r_0)}{2r^2}+\Lambda,
\end{equation}
where $r_0$ is an arbitrary constant. In the next section we shall see that, in the limit $\Q\to0$, equation~(\ref{MEq2}) can be reduced to (\ref{MEq1}), so that one can consider equation~(\ref{MEq2}) for any value of the electric charge $\Q$ without loss of generality. Thus, we will follow the second approach to inclusion of the electromagnetic field. This approach also reproduces the charged BTZ black hole \cite{Banados:1992wn} in the limit $\a_m\to0$.

\section{Description of the general metric and Hawking temperature}\label{sec:horizon}
Unlike the higher-dimensional case, when $D=3$, the arbitrary constant $r_{0}$ in (\ref{MEq2}) cannot be simply related with the black-hole mass. Therefore, as for the BTZ solution, we shall measure all dimensional quantities in units of the horizon radius $r_H$. Since $f(r_H)=0$, owing to (\ref{Lmetric}) we have $\psi(r_H)=\kappa r_H^{-2}$. Therefore, we find that
\begin{equation}\label{rdef}
\frac{\kappa}{r_H^2}+\sum_{m=2}^{\m}\frac{\a_m\kappa^{m}}{r_H^{2m}}=\frac{\Q^2\ln(r_H/r_0)}{2r_H^2}+\Lambda,
\end{equation}
what allows us to express $r_0$ in terms of the event horizon radius $r_H$.

Substituting ({\ref{rdef}}) into (\ref{MEq2}), we obtain the equation for $\psi(r)$ in terms of the event horizon radius,
\begin{eqnarray}\label{MEq3}
P[\psi(r)]&=&\frac{1}{r^2}\left(\kappa +\sum_{m=2}^{\m}\frac{\a_m\kappa^{m}}{r_H^{2m-2}}\right)
\\\nonumber&&
+\frac{\Q^2\ln(r/r_H)}{2r^2}+\Lambda\left(1-\frac{r_H^2}{r^2}\right).
\end{eqnarray}

Notice that if one substitutes the expression for mass in (\ref{MEq1}) in terms of $r_H$,
$$2\M= \frac{\kappa}{r_H^2} +\sum_{m=2}^{\m}\frac{\a_m\kappa^{m}}{r_H^{2m}}-\Lambda r_H^2+Q^2,$$
back into (\ref{MEq1}), then $Q^2$ cancels out and the corresponding equation for $\psi(r)$ coincides with (\ref{MEq3}) in the limit $\Q\to0$.
Therefore, we conclude that (\ref{MEq3}) or, equivalently, (\ref{MEq2}) describes the general static maximally symmetric solution in the three-dimensional Lovelock gravity  with the electric charge $\Q$.

Let us start from the assumption that there is a kind of three-dimensional analog of the cosmological horizon $r_C>r_H$, such that $f(r_C)=0$. Then, we have $\psi(r_C)=\kappa r_C^{-2}$, and one can express $\Lambda$ as follows:
\begin{equation}\label{LdSdef}
\Lambda=\sum_{m=2}^{\m}\a_m\kappa^m \frac{r_C^{2-2m}-r_H^{2-2m}}{r_C^2-r_H^2}-\frac{\Q^2\ln(r_C/r_H)}{2(r_C^2-r_H^2)}.
\end{equation}

If $\a_m=0$ then $\Lambda<0$, which means that the solution with positive $\Lambda$ cannot have a horizon. The interval $r_H<r<r_C$ corresponds to the region inside the inner horizon of the charged BTZ black hole, and the solution is asymptotically AdS.

However, from (\ref{LdSdef}) we see that when $\a_m\neq0$ the family of solutions to (\ref{MEq3}) can include asymptotically de Sitter black holes as well, with $\Lambda\to0$ as $r_C\to\infty$. The extreme value of $\Lambda$ is given by
\begin{equation}\label{LE}
\LE=\lim_{r_H\to r_C}\Lambda=-\sum_{m=2}^{\m}\a_m\kappa^m (m-1)r_H^{2-2m}-\frac{\Q^2}{4r_H^2}.
\end{equation}

In units of the event horizon we can also obtain a closed form for the Hawking temperature:
\begin{eqnarray}\label{Hawking}
T_H&=&\frac{f'(r_H)}{4\pi}=\frac{r_H(\LE-\Lambda)}{2\pi P'[\kappa r_H^{-2}]}
\\\nonumber&=&\frac{-4\Lambda r_H^2-\Q^2-4\sum_{m=2}^{\m}(m-1)\a_m\kappa^mr_H^{2-2m}}{8\pi r_H (1+\sum_{m=2}^{\m}m\a_m\kappa^{m-1}r_H^{2-2m})}\,.
\end{eqnarray}

When the denominator of (\ref{Hawking}) is positive, that is, $P'[\kappa r_H^{-2}]>0$, then the temperature decreases as the electric charge $\Q$ grows until its extreme value $\QE$, corresponding to $T_H=0$. Thus, the extreme charge $\QE$ is given by the relation
\begin{eqnarray}\label{extreme}
\QE^2&=&-4\Lambda r_H^2-4\sum_{m=2}^{\m}(m-1)\a_m\kappa^mr_H^{2-2m}
\\\nonumber&=&-4\Lambda r_H^2-4\a_2\frac{\kappa^2}{r_H^2}-8\a_3\frac{\kappa^3}{r_H^4}-12\a_4\frac{\kappa^4}{r_H^6}\ldots\geq0.
\end{eqnarray}

Inequality (\ref{extreme}) imposes the upper limit on $\Lambda$ for which the event horizon still exists:
\begin{equation}\label{Llimit}
\Lambda<-\sum_{m=2}^{\m}(m-1)\a_m\kappa^mr_H^{-2m}=-\frac{\a_2\kappa^2}{r_H^4}-\frac{2\a_3\kappa^3}{r_H^6}\ldots.
\end{equation}

Solutions of the field equations~(\ref{EEq}) and~(\ref{PEq}), satisfying the following inequality:
\begin{eqnarray}\label{perturbativecond}
P'[\kappa r_H^{-2}]&\equiv&1+\sum_{m=2}^{\m}m\a_m\kappa^{m-1}r_H^{2-2m}
\\\nonumber&&
=1+\frac{2\kappa\a_2}{r_H^2}+\frac{3\kappa^2\a_3}{r_H^4}+\frac{4\kappa^3\a_4}{r_H^6}+\ldots>0,
\end{eqnarray}
will here be called \emph{perturbative}, because in the limit $\a_m\to0$, they go over into the charged BTZ black hole ($\kappa=1$) or black string ($\kappa=0,-1$). Notice that inequality~(\ref{Llimit}) implies that the cosmological constant must be negative, if $\kappa=0$.

When $P'[\kappa r_H^{-2}]<0$, owing to equation~(\ref{Hawking}), $T_H$ grows when $\Q$ is increased, so that, if inequality~(\ref{Llimit}) is satisfied, then $\Q=\QE$ is the minimal charge, corresponding to $T_H=0$. Otherwise, if the value of $\Lambda$ is larger than the limit~(\ref{Llimit}), the uncharged black hole has nonzero Hawking temperature, and solutions can possess any electric charge.

Notice that, in the same way as for the higher-dimensional solutions \cite{Konoplya:2017lhs}, $P'[\psi(r)]$ cannot change its sign for $r\geq r_H$, because $\psi'(r)$ is divergent in the point $P'[\psi(r)]=0$, what leads to a singularity there. Therefore, if we study only regular black holes, we must choose values of $\a_m$ in such a way that $P'[\psi(r)]$ is either positive or negative for any $r\geq r_H$.

\begin{widetext}
\vspace{-16pt}

\section{Three-dimensional Gauss-Bonnet black hole}\label{sec:GaussBonnet}
In the regularized $3D$ Einstein-Gauss-Bonnet gravity~($\m=2$) equation~(\ref{MEq3}) has two solutions,
\begin{equation}\label{GBBH}
f(r)=\kappa-\frac{r^2}{2\a_2}\left(-1\pm\sqrt{1+2\a_2\frac{2\Lambda(r^2-r_H^2)+2\kappa+2\kappa^2\a_2/r_H^2+\Q^2\ln(r/r_H)}{r^2}}\right).
\end{equation}

\subsection{Perturbative branch}
The sign ``+'' corresponds to the perturbative branch, for which (\ref{perturbativecond}) reads
$P'[\kappa r_H^{-2}]=1+\dfrac{2\a_2\kappa}{r_H^2}>0.$

It is useful to rewrite this solution in the alternative form:
\begin{equation}\label{GBperturbative}
f(r)=\kappa-\frac{2\Lambda (r^2-r_H^2)+2\kappa+2\kappa^2\a_2/r_H^2+\Q^2\ln(r/r_H)}{1+\sqrt{1+2\a_2\left(2\Lambda r^2-2\Lambda r_H^2+2\kappa+2\kappa^2\a_2/r_H^2+\Q^2\ln(r/r_H)\right)/r^2}}.
\end{equation}
\end{widetext}

The constraint for the cosmological constant (\ref{Llimit}) has the form
$$\Lambda<-\frac{\a_2\kappa^2}{r_H^4},$$
implying that for $\a_2\geq0$ or $\kappa=0$ only the asymptotically AdS space allows for an event horizon. The electric charge $\Q$ must satisfy the inequality
$$\Q^2\leq\QE^2=-4\left(\Lambda r_H^2+\frac{\a_2\kappa^2}{r_H^2}\right)\,.$$

The metric function $f(r)$ has the following asymptotic:
\begin{equation}\label{aGBperturbative}
f(r)\to - \frac{2\Lambda}{1+\sqrt{1+4\a_2\Lambda}} r^2\,, \qquad r\to\infty\,,
\end{equation}
so that, considering the factor in front of $r^2$ as an effective cosmological constant, we can see that the latter has the same sign as $\Lambda$,
i.e., if $\Lambda<0$, the solution is asymptotically AdS, while for $\Lambda>0$, we have an asymptotically de Sitter black hole.

When $0\leq\Lambda<-\a_2\kappa^2r_H^{-4}$, using equation~(\ref{LdSdef}), we can express the cosmological constant in terms of the cosmological horizon $r_C$ as follows
\begin{equation}\label{GBLambda}
\Lambda=-\frac{\a_2\kappa^2}{r_H^2r_C^2}-\frac{\Q^2\ln(r_C/r_H)}{2(r_C^2-r_H^2)}.
\end{equation}

It is interesting to note that when $r_C\to\infty$ ($\Lambda\to0$), the solution is not asymptotically flat, unless $\Q=0$. The latter reads
\begin{eqnarray}\label{GBflat}
f(r)&=&\frac{-2\a_2\kappa^2/r_H^2-\kappa+\kappa\sqrt{1+\frac{4\a_2\kappa r_H^2+4\a_2^2\kappa^2}{r^2r_H^2}}}{1+\sqrt{1+\frac{4\a_2\kappa r_H^2+4\a_2^2\kappa^2}{r^2r_H^2}}}
\\\nonumber&=&-\frac{\a_2\kappa^2}{r_H^2}\left(1-\frac{(\a_2\kappa+r_H^2)^2}{r^2r_H^2}\right)+\O\left(\frac{1}{r^4}\right)\,.
\end{eqnarray}

This is a remarkable particular case of the general solution, when $\a_2<0$, representing the $(2+1)$-dimensional asymptotically flat black hole ($\kappa=1$) or string ($\kappa=-1$). When $\a_2 \to 0$, the metric function $f(r)$ vanishes and the solution does not exist, which agrees with the existence of only asymptotically AdS BTZ black holes at zero $\a_2$.

Equation~(\ref{aGBperturbative}) imposes an additional constraint: In order to have real solutions the Gauss-Bonnet parameter must obey
\begin{equation}\label{GBconstraint}
1+4\a_2\Lambda\geq0.
\end{equation}

When $\Lambda<0$ the condition (\ref{GBconstraint}) gives the upper bound for $\a_2\leq-1/4\Lambda$. When $\Lambda>0$, in addition to the lower bound for $\a_2\geq-1/4\Lambda$, there is a bound for the black-hole charge $\Q$, requiring that the solution be real.

When $\a_2$ is sufficiently small, the solution is real and can be thought as a Gauss-Bonnet corrected BTZ solution:
\begin{eqnarray}\label{BTZcorrection}
f(r)&=&-\Lambda(r^2-r_H^2)-\frac{\Q^2\ln(r/r_H)}{2}-\frac{\a_2\kappa^2}{r_H^2}
\\\nonumber&&
+\frac{\a_2}{r^2}\left(\kappa+(r^2-r_H^2)\Lambda+\frac{Q^2\ln(r/r_H)}{2}\right)^2+\O(\a_2^2)\,.
\end{eqnarray}

\subsection{Nonperturbative branch}
The sign ``-'' in (\ref{GBBH}) corresponds to the nonperturbative branch, for which
\begin{equation}\label{nonperturbativeineq}
1+\frac{2\a_2\kappa}{r_H^2}<0.
\end{equation}
These solutions correspond to small black holes, when $\a_2\kappa<0$, because $r_H$ must be smaller than $\sqrt{|2 \a_2\kappa|}$.

Since the metric function $f(r)$ has the following asymptotic
\begin{equation}
f(r)\to \frac{1}{\a_2}\frac{1+\sqrt{1+4\a_2\Lambda}}{2}r^2\,, \qquad r\to\infty\,,
\end{equation}
the effective cosmological constant and $\a_2$ have opposite signs.
In this case the value of $\Lambda$ can be considered as a small correction to the effective cosmological constant.

When $\a_2<-r_H^2/2$ ($\kappa=1$) the solution corresponds to an asymptotically de Sitter black hole. Using (\ref{GBLambda}), one can express $\a_2$ in terms of the cosmological horizon $r_C$ as follows
\begin{equation}\label{GBnonperturbativedS}
-\frac{\a_2}{r_H^2}=\Lambda r_C^2+\frac{\Q^2\ln(r_C/r_H)r_C^2}{2(r_C^2-r_H^2)}.
\end{equation}
We see that $\a_2\to-\infty$ as $r_C\to\infty$, so that inequality~(\ref{nonperturbativeineq}) holds. However, there is no black-hole solution in this limit, because the Hawking temperature
\begin{equation}\label{THnonperturbativedS}
T_H=-\frac{4\Lambda r_H^2+\Q^2+4\a_2/r_H^2}{8\pi r_H(1+2\a_2/r_H^2)}
\end{equation}
becomes negative. The reason for this is that the nonperturbative asymptotically de Sitter black hole has the minimal charge, given by (\ref{extreme}):
$$\Q^2\geq\QE^2=-4\Lambda r_H^2-4\a_2/r_H^2.$$
One cannot consider the limit $r_C\to\infty$ holding a constant value of $\QE$ or $\Lambda$ as well, because the expression in the square root in (\ref{GBBH}) becomes negative. We conclude, therefore, that the nonperturbative asymptotically de Sitter solution does not have a flat limit.

When $\a_2>r_H^2/2$ ($\kappa=-1$) the solution corresponds to the AdS black string. For small $\Lambda$ it reads
\begin{eqnarray}\label{AdSblackstring}
&&f(r)=-1+\dfrac{r^2}{2\a_2}\times
\\\nonumber&&
\phantom{f(r)=}\times\left(1+\sqrt{1-\dfrac{4\a_2}{r^2}+\dfrac{4\a_2^2}{r^2r_H^2}+\dfrac{2\a_2 Q^2\ln(r/r_H)}{r^2}}\right)
\\\nonumber&&
+\frac{\Lambda(r^2-r_H^2)}{\sqrt{1-\dfrac{4\a_2}{r^2}+\dfrac{4\a_2^2}{r^2r_H^2}+\dfrac{2\a_2 Q^2\ln(r/r_H)}{r^2}}} + \O(\Lambda^2)\,.
\end{eqnarray}

Finally we notice that, when $\kappa\a_2<0$, we cannot continuously decrease $r_H$ in order to go from the perturbative to the nonperturbative branch, because $T_H$ diverges in the limit $1+\kappa\a_2/r_H^2\to0$, which means that $r_H$ becomes a singular point of the solution.

\section{Third-order Lovelock gravity}\label{sec:Lovelock}
In the regularized third-order Lovelock gravity ($\m=3$) equation~(\ref{MEq3}) has generally three solutions. When $\a_3\geq\a_2^2/3$, only one solution is real:
\begin{equation}
f(r)=\kappa-\frac{\a_2r^2}{3\a_3}\left(A_+(r)-A_-(r)-1\right),
\end{equation}
where
\begin{eqnarray}\nonumber
A_{\pm}(r)&=&\sqrt[3]{\sqrt{F(r)^2+\left(\frac{3\a_3}{\a_2^2}-1\right)^3}\pm F(r)},\\\nonumber
F(r)&=&\frac{27\a_3^2}{2\a_2^3r^2}\Biggr(\Lambda r^2-\Lambda r_H^2+\kappa+\frac{\kappa^2\a_2}{r_H^2}+\frac{\kappa^3\a_3}{r_H^4}
\\\nonumber&&+\frac{Q^2\ln(r/r_H)}{2}\Biggr)+\frac{9\a_3}{2\a_2^2}-1\,.
\end{eqnarray}
In this case, for any $\psi(r)$ we have
$$P'(\psi(r))=1+2\a_2\psi(r)+3\a_3\psi^2(r)\geq(1+\a_2\psi(r))^2\geq0,$$
so that all solutions are perturbative.

The event horizon exists when (\ref{Llimit})
$$\Lambda<-\frac{\a_2\kappa^2}{r_H^4}-\frac{2\a_3\kappa^3}{r_H^6}.$$
Yet, the sign of the effective cosmological constant can be different from the sign of $\Lambda$. We notice that, when $\a_2<0$, the asymptotically de Sitter solutions always exist for sufficiently large black holes ($\kappa=1$).

The extreme charge is given by the relation (\ref{extreme}):
$$\Q^2\leq\QE^2=-4\left(\Lambda r_H^2+\a_2\frac{\kappa^2}{r_H^2}+2\a_3\frac{\kappa^3}{r_H^4}\right).$$

For $\a_2<\a_2^2/3$, there are three real solutions to equation~(\ref{MEq3}). In principle, for each set of the values of $r_H$, $\a_2$, and $\a_3$, the solution can be given  in a closed, but cumbersome form. Yet, such an analysis, as well as the analysis of solutions in higher-order Lovelock gravity, is beyond the scope of the present paper. We believe that, for practical purposes, in higher-order Lovelock gravity it is easier to work with numerical solutions of (\ref{MEq3}) rather than to derive the lengthy expressions with various branches in their closed forms.

\section{Conclusions}
A black hole in the $D>4$ Einstein-Gauss-Bonnet gravity and its Lovelock generalization were extensively studied and a number of interesting properties were observed. For example, the lifetime of only a slightly Gauss-Bonnet corrected black hole is characterized by a few orders longer lifetime and a smaller evaporation rate \cite{Konoplya:2010vz}. The eikonal quasinormal modes break down the correspondence between the eikonal quasinormal modes and null geodesics \cite{Cardoso:2008bp,Konoplya:2017wot}. Apparently, one of the interesting properties of higher curvature corrected black holes is the gravitational instability: When the coupling constants are not small enough, the black holes are unstable and the instability develops at high multipole numbers   \cite{Dotti:2005sq,Gleiser:2005ra,Konoplya:2017lhs,Konoplya:2017zwo,Yoshida:2015vua,Takahashi:2011qda,Konoplya:2008ix,Cuyubamba:2016cug,Takahashi:2012np}.

In higher dimensions, as well as when considering $D=4$ Einstein-Gauss-Bonnet black holes coupled to a dilaton or other scalar field, all the effects due to the higher curvature corrections are analyzed numerically (see, for example, \cite{Takahashi:2010,Maselli:2014fca,Nampalliwar:2018iru,Konoplya:2019hml,Xu:2019krv,Zinhailo:2019rwd,Mishra:2019ged,Churilova:2019sah,Cuyubamba:2019qtz,Blazquez-Salcedo:2020rhf} and references therein). The model of the three-dimensional Gauss-Bonnet and Lovelock corrected black holes and strings considered in this paper could be a much simpler model for analysis of various effects in the black-hole background, which could, possibly, be analyzed analytically and, thereby, give a clearer understanding of various phenomena in the presence of the higher curvature corrections. The existence of the asymptotical flat black-hole solution in the $(2+1)$-dimensional spacetime gives addition advantages for this. One of the possible nearest future aims could be the analysis of the quasinormal spectra \cite{Konoplya:2011qq} of the above black holes.

It is interesting to notice that the black brane solution (\ref{GBperturbative}) for $\kappa=0$ obtained in our work as a result of the dimensional regularization was also reproduced in an alternative approach proposed recently in \cite{Fernandes:2020nbq,Hennigar:2020lsl}, where there is a well-defined action,
\begin{eqnarray}\nonumber
S&=&\int d^3 x \sqrt{-g}\Bigl[R-2\Lambda-2\a_2(g^{\mu\nu}\partial_{\mu} \phi \partial_{\nu} \phi)^2
\\\nonumber&&
-2\a_2(2R^{\mu\nu}-(R+2\Box \phi)g^{\mu\nu})\partial_{\mu} \phi \partial_{\nu} \phi
\\\nonumber&&
-\a_2\phi (R_{\mu\nu\lambda\sigma}R^{\mu\nu\lambda\sigma}-4 R_{\mu\nu}R^{\mu\nu}+R^2)\Bigr]\,.
\end{eqnarray}
Thus, the black brane metric~(\ref{GBperturbative}) is also an exact solution of the scalar-tensor theory with $\phi=\ln(r/l)$ \cite{Hennigar:2020fkv}. This means that although it is not clear whether the dimensional regularization leads to any well-defined theory in (2+1) dimensions, it can be an effective tool for obtaining exact solutions. Unlike black brane solution~(\ref{GBperturbative}), the black-hole solutions obtained in this work have not been yet reproduced in some well-defined theory, and this raises another appealing question: whether the scalar-tensor theory used in \cite{Hennigar:2020fkv} could be modified in such a way that the $\kappa=\pm1$ solutions are allowed as well.

\bigskip\bigskip

\acknowledgments
The authors acknowledge the support of Grant No.~19-03950S of Czech Science Foundation (GAČR). This publication has been prepared with partial support of the ``RUDN University Program 5-100'' (R. K.).

\end{document}